\documentclass[twocolumn,showpacs,amsmath,amssymb,superscriptaddress]{revtex4-2}
\usepackage{dcolumn,braket,color,amsmath,footnote,hyperref,bm}
\usepackage{graphicx,bm,url,longtable}
\usepackage{tabularx,multirow,braket}
\usepackage{fancybox}
\usepackage{tikz}
\usetikzlibrary{matrix}

\newcommand\+{\dagger}
\newcommand{\omin}{3,\mathrm{min}}
\newcommand{\qmin}{2,\mathrm{min}}

\begin{document}

\title{Quadrupole-octupole coupling and the onset of octupole deformation in actinides}

\author{K.~Nomura}
\email{knomura@phy.hr}
\affiliation{Department of Physics, Faculty of Science, University of Zagreb, HR-10000, Croatia}

\author{R.~Rodr\'iguez-Guzm\'an}
\affiliation{Physics Department, Kuwait University, 13060 Kuwait, Kuwait}

\author{L.~M.~Robledo}
\affiliation{Departamento de F\'\i sica Te\'orica and CIAFF, Universidad
Aut\'onoma de Madrid, E-28049 Madrid, Spain}

\affiliation{Center for Computational Simulation,
Universidad Polit\'ecnica de Madrid,
Campus de Montegancedo, Bohadilla del Monte, E-28660-Madrid, Spain
}

\author{J.~E.~Garc\'ia-Ramos}
\affiliation{Departamento de Ciencias Integradas y Centro de Estudios 
Avanzados en F\'isica, Matem\'atica y Computaci\'on, Universidad de Huelva, 
E-21071 Huelva, Spain}

\affiliation{Instituto Carlos I de F\'{\i}sica Te\'orica y Computacional,  
Universidad de Granada, Fuentenueva s/n, 18071 Granada, Spain}

\date{\today}

\begin{abstract}
The evolution of quadrupole and octupole collectivity and their 
coupling is investigated in a series of even-even  isotopes of the actinide
Ra, Th, U, Pu, Cm, and Cf with neutron number in the interval 
$130\leqslant N\leqslant 150$. The Hartree-Fock-Bogoliubov 
approximation, based on the parametrization D1M of the Gogny energy 
density functional, is employed  to generate potential energy surfaces 
depending upon the axially-symmetric quadrupole and octupole shape 
degrees of freedom. The mean-field energy surface is then mapped onto 
the expectation value of the $sdf$ interacting-boson-model Hamiltonian 
in the boson condensate state as to determine the strength parameters 
of the boson Hamiltonian. Spectroscopic properties related to the 
octupole degree of freedom are produced by diagonalizing the mapped 
Hamiltonian. Calculated low-energy negative-parity spectra, 
$B(E3;3^{-}_{1}\to 0^{+}_{1})$ reduced transition rates, and effective 
octupole deformation suggest that the transition from nearly spherical 
to stable octupole-deformed, and to octupole vibrational states occurs 
systematically in the actinide region. 
\end{abstract}

\maketitle

\section{Introduction}

Over the decades, octupole deformation in nuclei and the related 
spectroscopy of negative-parity collective states has been an active 
research field in low-energy nuclear physics 
\cite{butler1996,butler2016}. It is well known that the ground 
state of most medium-mass and heavy nuclei is reflection symmetric and 
therefore the dominant intrinsic deformation is of quadrupole 
character. On the other hand, there are a handful of nuclear systems 
where reflection symmetry is broken, giving rise to an 
octupole-deformed ground state. The octupole shape is expected to be 
present in those mass regions corresponding to protons $Z$ and neutron 
numbers $N$ close to 34, 56, 88, and 134 \cite{butler1996,butler2016}. 
Observables characteristic of the ground state static octupole 
deformation are low-lying negative-parity bands, which form well 
deformed quadrupole deformed nuclei and approximate alternating-parity 
doublet with the ground-state positive-parity band, and enhanced 
electric dipole and octupole transition rates. 
%In particular, non-zero atomic electric dipole moment would implies
%violation of CP symmetry, and this represents extensions of the
%Standard Model of elementary particles \cite{engel2013}. 
Fingerprints of stable octupole shapes have been found experimentally in
light actinides ($^{220}$Rn, $^{224}$Ra and $^{222,228}$Ra 
\cite{gaffney2013,butler2020a} and $^{228}$Th \cite{chishti2020}) and
lanthanides ($^{144,146}$Ba \cite{bucher2016,bucher2017}). 
Within this context, numerous theoretical investigations
have been made to predict, and support evidence of octupole deformation
by means of various theoretical models: 
macroscopic-microscopic models \cite{naza1984b,leander1985,moeller2008},
self-consistent mean-field (SCMF) methods 
based on the nuclear density functional theory 
\cite{MARCOS1983,BONCHE1986,BONCHE1991,heenen1994,ROBLEDO1987,ROBLEDO1988,EGIDO1990,EGIDO1991,EGIDO1992,GARROTE1998,GARROTE1999,long2004,robledo2010,robledo2011,erler2012,robledo2012,rayner2012,robledo2013,robledo2015,bernard2016,agbemava2016,agbemava2017,xu2017,xia2017,ebata2017,rayner2020,cao2020,rayner2020oct},
interacting boson model (IBM)
\cite{engel1987,zamfir2001,zamfir2003,nomura2013oct,nomura2014,nomura2015,nomura2020oct},
geometrical collective models \cite{bonatsos2005,lenis2006,bizzeti2013},
and cluster models \cite{shneidman2002,shneidman2003}.

Octupole collective excitations in the light actinide nuclei with 
$Z\approx 88$ and $N\approx 134$ have been extensively studied both 
experimentally and theoretically. However, spectroscopic data for those 
actinide nuclei heavier than Th, such as Pu, Cm, and Cf isotopes, are 
scarce, especially in the neutron-deficient side of the nuclear chart 
with $N\approx 134$. This is because these isotopes are close to the 
proton drip-line and have not been accessible so far by experiments. It 
is, nevertheless, worth to explore theoretically whether the robustness 
of the neutron octupole magic number $N=134$ in actinides holds when 
one departs from $Z\approx 88$ towards the proton drip-line.

In the present work, we employ the EDF-to-IBM
mapping procedure \cite{nomura2008} for a theoretical calculation of the properties of
octupole collective excitations. This procedure involves two main steps: first, for each nucleus a potential
energy surface (PES) depending upon the axially-symmetric quadrupole
$\beta_{2}$ and octupole $\beta_{3}$ shape degrees of freedom is computed
within the constrained SCMF method with a choice of an universal energy
density functional and pairing force. In a second step, the PES is mapped
onto the expectation value of the bosonic Hamiltonian in the 
condensate state of the monopole $s$, quadrupole 
$d$, and octupole $f$ bosons in order to fix some of the model's parameters \cite{IBM,engel1987}.
Subsequent diagonalization of the resulting boson Hamiltonian yields excitation
energy spectra and transition strengths. 
%Diagonalization of the resulting boson Hamiltonian yields excitation
%spectra and transition strengths. 
%For the details about the mapping
%procedure for octupole deformed nuclei, the reader is referred to
%Refs.~\cite{nomura2013oct,nomura2014}. 
The mapping procedure has been initially implemented
\cite{nomura2013oct,nomura2014} in the study of spectroscopic properties of
reflection-asymmetric Ba, Sm, Ra, and Th 
nuclei using the relativistic density-dependent point-coupling
(DD-PC1) EDF \cite{DDPC1} to generate the microscopic PES. 
More recently, updated calculations have been performed
\cite{nomura2020oct} in the $^{218-238}$Ra and $^{220-240}$Th isotopic chains within the
mapped IBM framework based on the Gogny D1M EDF \cite{D1M}. 
In those studies, consistent with the empirical trend, we 
have identified a most pronounced octupolarity around the 
neutron number $N=134$ in the systematic of calculated physical
observables. 
%and that principal results are essentially the
%same relardless of whether relativistic or non-relativisitic EDF is
%employed. 

In view of the renewed interest in a global search for the static octupole
deformation beyond Ra and Th isotopes, here we extend the analysis of
\cite{nomura2020oct} to heavier actinide nuclei $^{222-242}$U,
$^{224-244}$Pu, $^{226-246}$Cm, and $^{228-248}$Cf, and verify whether
the octupole-related shape phase transition generally occurs in the
actinide region, that is, the onset of stable 
octupole deformation at $N\approx 134$ and octupole softness starting
from $N\approx 138$. 
With the present study we also  aim to assess the performance of the 
EDF-based IBM approach in the global description of octupole collective
states in the actinide region. The results further include quantitative
predictions on spectroscopy in proton-rich actinides that
have not been explored so far by experiment. This has not been
possible in the previous IBM calculations, since they are mostly fit to known
experimental data. Hence, the present work points not only to an alternative
EDF-based approach to the detailed spectroscopy of a large number of
actinide nuclei, but it is a first implementation of the IBM framework in the
octupole-related spectroscopic studies on proton-rich heavy actinides,
that is based on the microscopic EDF.

It should be noted that interplay between quadrupole and octupole 
degrees of freedom in the low-lying negative parity collective states 
of U, Pu, Cm, and Cf nuclei has also been studied within the framework 
of the generator coordinate method (GCM) 
\cite{RS,bender2003,robledo2019} using the Gogny D1M EDF 
\cite{rayner2020oct}. The GCM calculation is, however, quite time 
consuming especially for heavy systems or when the number of collective 
coordinates increases. Computational complexity also prevents to have 
access to high spin states like, for instance, the members of 
alternating parity rotational bands. In Refs.~\cite{xia2017,xu2017}, a 
large number of heavy and superheavy nuclei up to No isotopes with mass 
$A\approx 300$ have been analyzed by solving quadrupole-octupole 
collective Hamiltonian, with parameters specified by the relativistic 
EDF calculations. Detailed spectroscopy of stable U and Pu nuclei has been 
explored within a purely phenomenological $spdf$-IBM framework in 
Refs.~\cite{zamfir2003,spieker2018}.

The paper is organized as follows. The mapping procedure used to obtain 
the IBM Hamiltonian is illustrated in Sec.~\ref{sec:method}. The 
results of our analysis are discussed in Sec.~\ref{sec:results}, 
including the Gogny-D1M quadrupole-octupole SCMF-PESs, i.e., the 
microscopic building blocks of the calculations, mapped IBM-PESs, 
low-energy excitation spectra, transition properties, and the effective 
$\beta_{2}$ and $\beta_{3}$ deformation parameters. Finally, 
Sec.~\ref{sec:summary} is devoted to the concluding remarks and work 
perspectives.

% ----------------------------------------------------------------------
%  Section:                                        Theoretical method
%
% ----------------------------------------------------------------------

\section{Theoretical method \label{sec:method}}

To obtain the quadrupole-octupole SCMF-PESs, the HFB equation has been
solved with constrains on the axially symmetric quadrupole
$\hat{Q}_{20}$ and octupole $\hat{Q}_{30}$ operators
\cite{rayner2012,rayner2020oct}. 
%\begin{align}
%&\hat{Q}_{20} = z^{2} - \frac{1}{2}\left(x^{2} + y^{2}  \right) 
%\nonumber\\
%&\hat{Q}_{30} = z^{3} - \frac{3}{2} z\left(x^{2} + y^{2}  \right)
%\end{align}
The mean values $\langle \Phi_\mathrm{HFB} |\hat{Q}_{20}| \Phi_\mathrm{HFB} \rangle = Q_{20}$
and $\langle \Phi_\mathrm{HFB} |\hat{Q}_{30}| \Phi_\mathrm{HFB} \rangle
= Q_{30}$ define the quadrupole and octupole deformation parameters
$\beta_{\lambda}$ ($\lambda=2,3$), i.e.,  $\beta_{\lambda} =
\sqrt{4 \pi (2\lambda +1)}Q_{\lambda 0}/(3 R_{0}^{\lambda} A)$,  
with $R_0=1.2 A^{1/3}$ fm. The constrained  calculations 
provide a set of HFB states $| \Phi_\mathrm{HFB} (\beta_{2},\beta_{3})\rangle$ 
labeled by their  
static deformation parameters $\beta_{2}$ and $\beta_{3}$.
The HFB energies $E_\mathrm{HFB}(\beta_{2},\beta_{3})$ associated
with those HFB states define the so-called  
SCMF-PESs used in this work. As the HFB energies satisfy the 
property $E_\mathrm{HFB}(\beta_{2},\beta_{3}) = E_\mathrm{HFB}(\beta_{2},-\beta_{3})$
only positive $\beta_{3}$ values
are considered when plotting  the SCMF-PESs.

Excitation energies and transition probabilities of the quadrupole and
octupole collective states are computed by diagonalizing the IBM
Hamiltonian that is determined with
microscopic input from the Gogny-HFB SCMF calculation - see below. 
For computing negative-parity states, we consider the  $J=0^+$ ($s$), $2^+$ ($d$), and $J=3^-$ ($f$)
bosons  as building blocks of
the IBM. The total number of bosons $n=n_{s}+n_{d}+n_{f}$ is conserved for 
a given nucleus, and is equal to half the number of valence
nucleons. The doubly-magic nucleus $^{208}$Pb is taken here as the inert
core, and thus $n=(A-208)/2$ for a nucleus with mass $A$. 
We adopt the $sdf$-IBM Hamiltonian
\cite{nomura2020oct}:
\begin{align}
\label{eq:ham}
\hat H_{\mathrm{IBM}} = 
\epsilon_d\hat n_{d} + \epsilon_{f}\hat{n}_{f} 
+ \kappa_{2}\hat{Q}_{2}\cdot\hat{Q}_{2} + \rho\hat{L}\cdot\hat{L} 
+ \kappa_{3}\hat{Q}_{3}\cdot\hat{Q}_{3}. 
\end{align}
The first (second) term represents the number operator for
the $d$ ($f$) bosons with $\epsilon_{d}$ ($\epsilon_{f}$) being the single
$d$ ($f$) boson energy relative to the $s$ boson one. 
The third, fourth, and fifth terms represent the quadrupole-quadrupole
interaction, the rotational term, and the octupole-octupole 
interaction, respectively. 
The quadrupole $\hat{Q}_{2}$, the angular momentum $\hat{L}$, and the
octupole $\hat{Q}_{3}$
operators are expressed as
\begin{align}
\label{eq:q3}
& \hat Q_{2}=s^{\dagger}\tilde d+d^{\dagger}\tilde s+\chi_{d}[d^{\dagger}\times\tilde
  d]^{(2)}+\chi_{f}[f^{\dagger}\times\tilde f]^{(2)} \\
& \hat
L=\sqrt{10}[d^{\dagger}\times\tilde{d}]^{(1)}+\sqrt{28}[f^\+\times\tilde
f]^{(1)} \\
&\hat{Q}_{3}=s^{\dagger}\tilde{f}+f^{\dagger}\tilde
  s+\chi_{3}[d^{\dagger}\times\tilde
  f+f^{\dagger}\times\tilde
  d]^{(3)}. 
\end{align}
Note that the term proportional to
$(d^{\+}\tilde{d})^{(1)}\cdot (f^{\+}\tilde{f})^{(1)}$ in the
$\hat{L}\cdot\hat{L}$ term has been neglected \cite{nomura2020oct}. 
The parameters 
$\epsilon_d$, $\epsilon_f$, $\kappa_{2}$, $\rho$, $\chi_{d}$, $\chi_{f}$, 
$\kappa_{3}$, and $\chi_{3}$ of the $sdf$-IBM Hamiltonian 
are determined, for each nucleus, in such a way  
\cite{nomura2015,nomura2020oct} that 
the expectation value of the $sdf$-IBM Hamiltonian in the boson
condensate state,
$E_\mathrm{IBM}(\beta_2,\beta_3)=
\bra{\phi(\beta_{2},\beta_{3})}\hat{H}_\mathrm{IBM}\ket{\phi(\beta_{2},\beta_{3})}$, 
reproduces the Gogny-HFB SCMF PES $E_\mathrm{HFB}(\beta_2,\beta_3)$ in
the neighborhood of the global minimum. 
The boson condensate wave function is given by 
\cite{ginocchio1980}: 
\begin{align}
 \label{eq:coherent}
|\phi(\beta_{2},\beta_{3})\rangle=
(n!)^{-1/2}(s^\+ +\bar\beta_{2}d_0^\+ + \bar\beta_{3}f_{0}^\+)^{n}
\ket{0}, 
\end{align}
where $\ket{0}$ denotes the inert core, i.e., $^{208}$Pb. 
The amplitudes $\bar\beta_{2}$ and $\bar\beta_{3}$ entering the 
definition of the boson condensate wave function are proportional 
to the deformation parameters $\beta_{2}$ and $\beta_{3}$ of the
fermionic space, $\bar\beta_{2}=C_2\beta_2$ and
$\bar\beta_{3}=C_3\beta_3$ 
\cite{ginocchio1980,nomura2014,nomura2015}, with dimensionless proportionality 
constants  $C_2$ and $C_3$. Their values are also determined by the
mapping procedure so that the location of the global minimum in the SCMF-PES,  
denoted by $\beta_{\qmin}$ and $\beta_{\omin}$, is reproduced. 
A more detailed description of the whole procedure can be found in
Ref.~\cite{nomura2020oct}. As for the analytical expression of the IBM-PES
$E_\mathrm{IBM}(\beta_2,\beta_3)$, we refer the reader to
Ref.~\cite{nomura2015}. 

For the numerical diagonalization of the Hamiltonian $\hat{H}_\mathrm{IBM}$
(\ref{eq:ham}), the computer code \textsc{ArbModel} \cite{arbmodel} has been
used. 
%However, the current implementation of the code is restricted to the
%diagonalization of an $sdf$-IBM Hamiltonian with boson number $n=16$,
%i.e., $A=240$. Hence, for those nuclei with 
%$A>240$, we have used the \textsc{arbmodel} code \cite{arbmodel}. 
%and considered one $f$ boson in the model space. 

%-----------------------------------------------------------------------
%
% 	PESs for Pu
%
%-----------------------------------------------------------------------
\begin{figure}[htb!]
\begin{center}
\includegraphics[width=\linewidth]{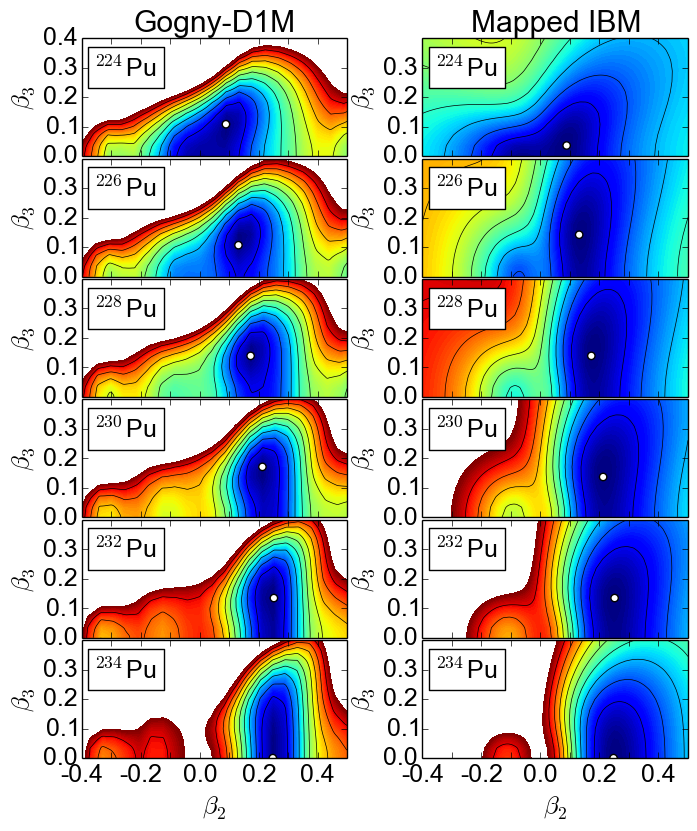}
\caption{SCMF-PESs computed with the Gogny-D1M EDF for 
the nuclei $^{224-234}$Pu (left), and the corresponding mapped
 $sdf$-IBM PESs (right). 
The color code indicates the total HFB and IBM energies in MeV units, 
plotted up to 10 MeV with respect to the global minimum. Energy
 difference between neighboring contours is 1 MeV. For each
 nucleus, the global minimum is
 indicated by an open circle.} 
\label{fig:pes-pu}
\end{center}
\end{figure}

%-----------------------------------------------------------------------
%
% 	PESs for Cf
%
%-----------------------------------------------------------------------
\begin{figure}[htb!]
\begin{center}
\includegraphics[width=\linewidth]{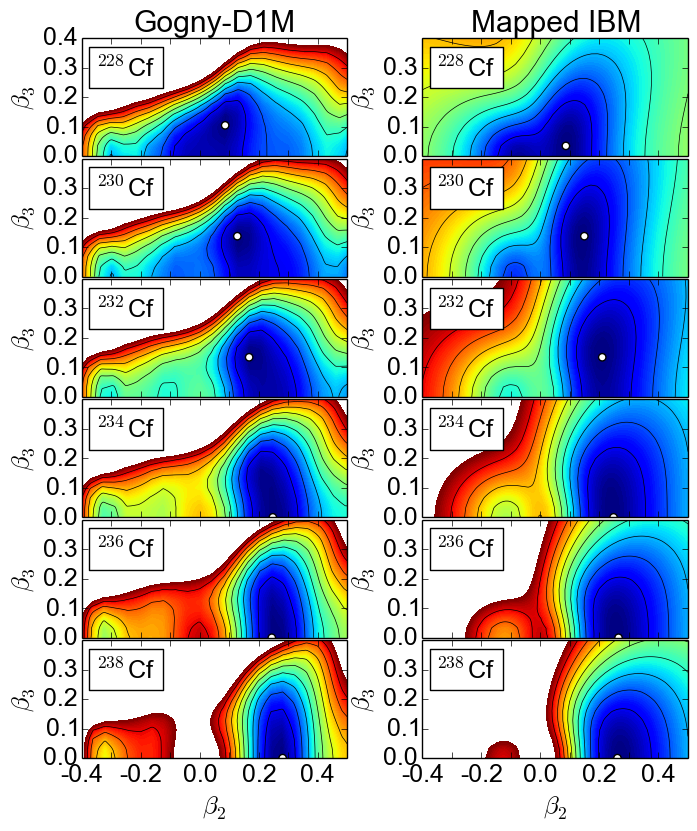}
\caption{Same as for Fig.~\ref{fig:pes-pu}, but for $^{248-238}$Cf.} 
\label{fig:pes-cf}
\end{center}
\end{figure}

%-----------------------------------------------------------------------
%
% 	E_oct, etc
%
%-----------------------------------------------------------------------
\begin{figure}[htb!]
\begin{center}
\includegraphics[width=\linewidth]{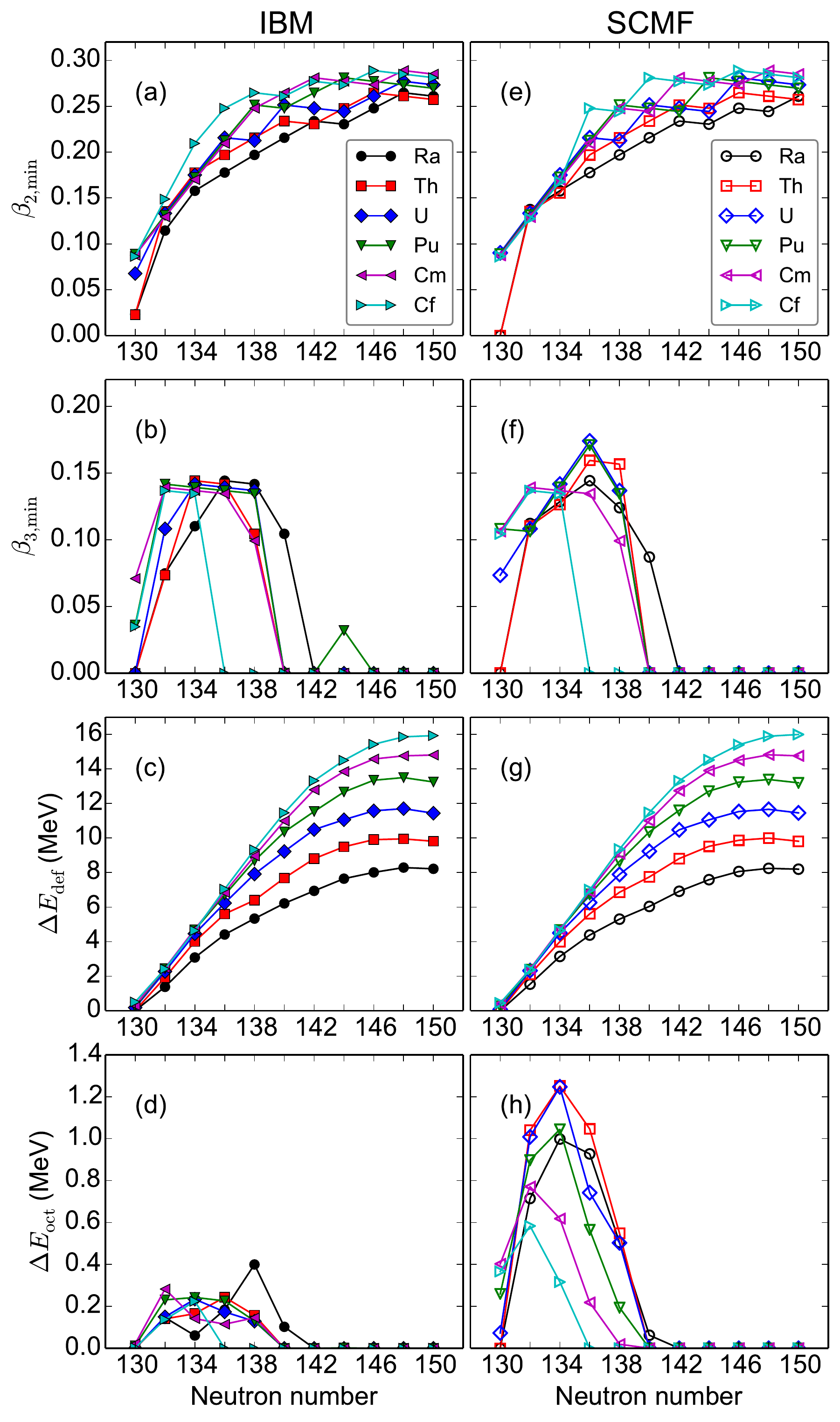}
\caption{The $\beta_{2,\mathrm{min}}$ and $\beta_{3,\mathrm{min}}$
values, corresponding to the ground state minimum are plotted as a function
of neutron number in panels (a) and (b), respectively. In panel (c) the
deformation energy  
$\Delta E_{\mathrm{def}}$, defined as the energy difference between the global
 minimum and the spherical configuration is plotted as a function of neutron 
 number. The octupole
deformation energy $\Delta E_{\mathrm{oct}}$, defined as the energy
difference between the global minimum and the quadrupole deformed
minimum along the $\beta_{3}=0$ axis, for the mapped
IBM-PESs is plotted in panel (d). The corresponding quantities for the SCMF-PESs are also
plotted in the right hand side panels from  (e) to (h). The results for the Ra and
Th nuclei have been taken from Ref.~\cite{nomura2020oct}. 
See the main text for definitions of the above quantities.}
\label{fig:eoct}
\end{center}
\end{figure}

%-----------------------------------------------------------------------
%
% 	PARAMETERS
%
%-----------------------------------------------------------------------
\begin{figure}[htb!]
\begin{center}
\includegraphics[width=\linewidth]{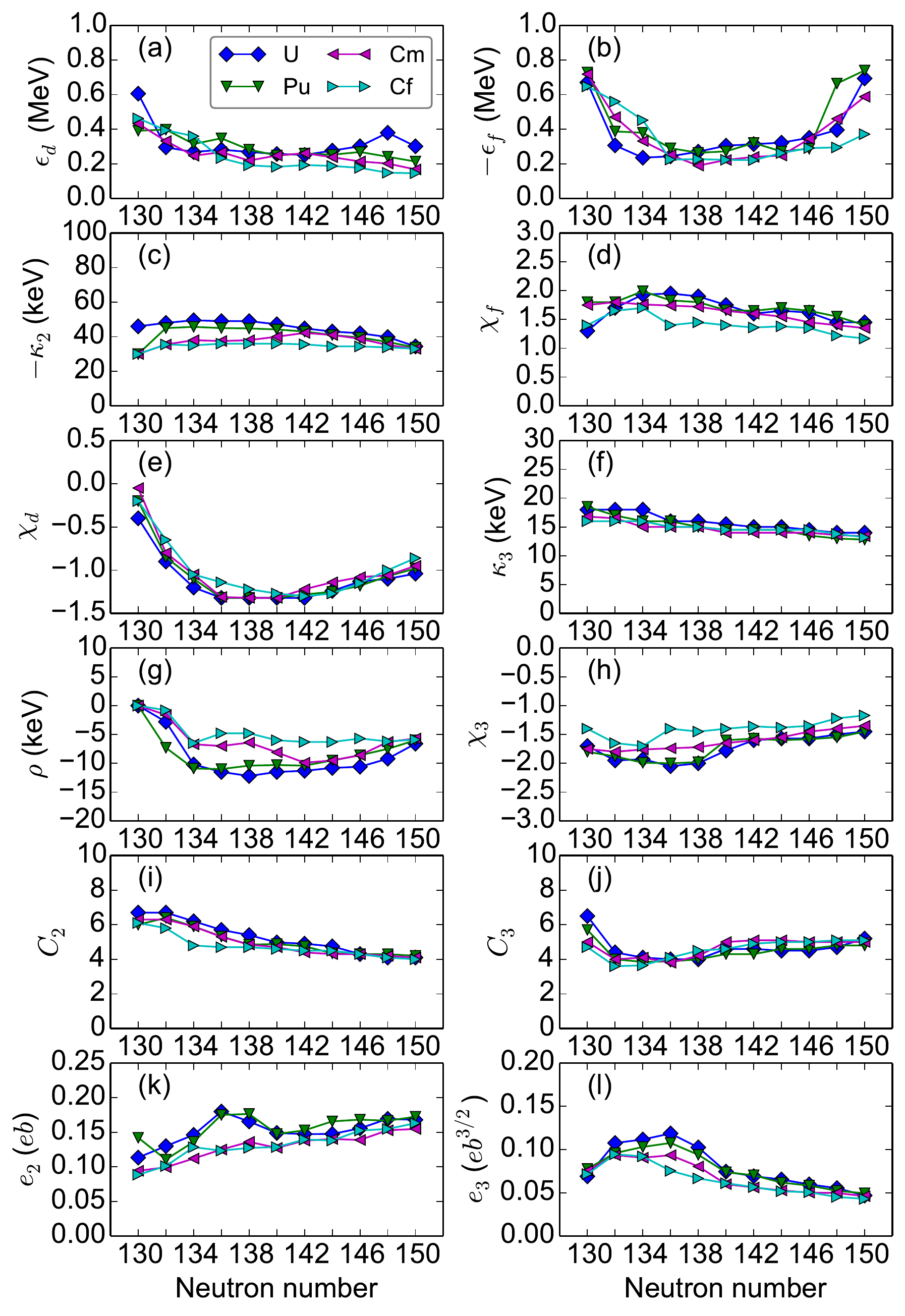}
\caption{The strength parameters $\epsilon_d$, $\epsilon_f$, $\kappa_2$,   $\chi_{f}$, $\chi_{d}$,
$\kappa_{3}$, $\rho$ and $\chi_{3}$ of the $sdf$-IBM 
Hamiltonian Eq. (\ref{eq:ham}) are plotted as a function of neutron number in panels
from (a) to (h) for the four isotopic chains considered. The coefficients $C_{2}$ and $C_{3}$ 
connecting the IBM and microscopic
quadrupole and octupole deformation parameters are plotted in panels (i) and (j). Finally, 
the boson effective charges for the quadrupole $e_{2}$ and octupole $e_{3}$
transitions are plotted as a
function of the neutron number in panels (k) and (l), respectively. The
 parameters for Ra and Th isotopes can be found in Ref.~\cite{nomura2020oct}.}
\label{fig:parameter}
\end{center}
\end{figure}

% ----------------------------------------------------------------------
%  Section:                                      Results and discussions
%
% ----------------------------------------------------------------------

\section{Results and discussions\label{sec:results}}

% Subsection:                                       Gogny-D1M SCMF-PESs

\subsection{Gogny-D1M SCMF-PESs\label{sec:pes}}

As representative cases, the Gogny-D1M SCMF-PESs for $^{224-234}$Pu
and $^{228-238}$Cf are depicted in the left columns of
Figs.~\ref{fig:pes-pu} and \ref{fig:pes-cf}, respectively. 
In most of the Pu isotopes, a non-zero $\beta_{3}$ minimum
$\beta_{3,\mathrm{min}}$ is found in the 
interval of $0.1\leqslant\beta_{3}\leqslant 0.2$. 
The most pronounced octupole minimum is obtained at $N=134$
($^{228}$Pu) around $\beta_{\omin}\approx 0.17$. 
For Cf nuclei in Fig.~\ref{fig:pes-cf}, only three isotopes 
are octupole deformed with $\beta_{\omin}<0.15$. 
Note that for those isotopes with $N>140$, the potential energy becomes softer in
$\beta_{3}$ and no octupole deformation is found. 

The systematic of the SCMF-PESs for U and Cm isotopes is similar to the
one for Pu and Cf isotopes, and the PESs for Ra and Th can be found in
Ref.~\cite{nomura2020oct}. It is also worth mentioning that   
the Gogny-D1M SCMF-PESs used in this paper are very similar to 
those obtained in Ref.~\cite{robledo2012} using the Gogny-D1S \cite{D1S} and
D1N \cite{D1N}, and the Barcelona-Catania-Paris
(BCP) \cite{robledo2010} EDFs. 
Recent comparisons of several non-relativistic Skyrme and relativistic EDFs in a survey of octupole
correlations can be found in Refs.~\cite{cao2020} and
\cite{agbemava2016}, respectively. In those references, it is noticed that 
for most of the adopted EDFs
pronounced octupole mean-field minimum occur around $N=134$. 

% Subsection:                                           Mapped IBM-PESs 

\subsection{Mapped IBM-PESs\label{sec:IBMPES}}

The corresponding IBM-PESs are drawn on the right-hand sides of
Figs.~\ref{fig:pes-pu} and \ref{fig:pes-cf}. One can clearly observe the similarities
between the SCMF and IBM PESs: the topography of the IBM-PES 
changes with the neutron number in a similar way as the SCMF-PES, from nearly
spherical configurations ($N\approx 130$) to pronounced octupole deformation
($N\approx 134$) continuing with $\beta_{3}$-soft ($N\approx 138$). 
The  IBM-PESs are generally softer  than the SCMF-PESs. This is related to
the more restricted configuration space of the IBM as compared to the one of the
SCMF model.

In Fig~\ref{fig:eoct} we show, as a function of neutron number,  
the values of several quantities characterizing the self-consistent
minima found in the IBM (left panels) and SCMF PES (right panels). 
On panels (a) and (b) the values of the quadrupole and octupole deformation parameters of
the absolute minimum $\beta_{\qmin}$ and
$\beta_{\omin}$ are plotted. On panel (c) the quadrupole deformation energy $\Delta E_\mathrm{def}$ 
defined as the energy difference between the energy at the
global minimum and at the spherical point: 
\begin{align}
%\Delta E_\mathrm{def}=E_\mathrm{IBM}(\beta_{\qmin},\beta_{\omin})-E_\mathrm{IBM}(0,0),
\Delta E_\mathrm{def}=E(\beta_{\qmin},\beta_{\omin})-E(0,0),
\end{align}
is shown, where $E(\beta_{2},\beta_{3})$ refers either to the SCMF-PES
or the IBM-PES. Finally, the octupole deformation energies $\Delta
E_\mathrm{oct}$, 
defined as the energy difference between the energy at the global
minimum and the local minimum on the $\beta_{3}=0$ axis, i.e.
\begin{align}
% \Delta E_\mathrm{oct} = E_\mathrm{IBM}(\beta_{\qmin},\beta_{\omin}) - E_\mathrm{IBM}(\beta_{\qmin}',0),
 \Delta E_\mathrm{oct} = E(\beta_{\qmin},\beta_{\omin}) - E(\beta_{\qmin}',0),
\end{align}
is shown in panel (d). In the above expression 
$\beta_{\qmin}'$ stands for the $\beta_{2}$ deformation parameter corresponding to
the local minimum on the $\beta_{3}=0$ axis. 
The corresponding quantities calculated with the SCMF-PESs
$E_\mathrm{HFB}(\beta_{2},\beta_{3})$ are included in panels from (e) to
(h) in the plot. The systematic as a function of neutron number
of most of these quantities is basically the same when looking at  the
SCMF and IBM PESs. 

Only one exception can be seen in the quantity $\Delta
E_\mathrm{oct}$ shown in Fig.~\ref{fig:eoct} panel (d): the IBM values
are a factor between two to five lower than the SCMF ones. 
This discrepancy reflects the fact that the IBM-PESs are much softer in
$\beta_{3}$ direction than the SCMF-PESs: the latter displays a much steeper
potential in $\beta_{3}$ than the former (see also Figs.~\ref{fig:pes-pu} and
\ref{fig:pes-cf}). 
It is partly attributed to the fact that the analytical
form of the IBM-PES \cite{nomura2015} is so restricted, comprising only
limited number and species of bosons, that it is not able to reproduce 
in full detail the topology of
the SCMF-PES, which is much steeper in $\beta_{3}$ direction, but only
the overall topology of the SCMF-PES typically up 
to a few MeV from the minimum. However, the SCMF
solutions within this energy range are most relevant to low-lying
states. In addition, the fact that the IBM-PESs are
considerably $\beta_{3}$ soft as compared to the SCMF ones presents a
general feature of the IBM framework, but is considered to be of minor
relevance to reproducing spectroscopic
properties of low-energy yrast state and thus does not alter the main
conclusions.

% Subsection:                                     Derived IBM parameters

\subsection{Derived IBM parameters\label{sec:para}}

In Fig.~\ref{fig:parameter} we display the IBM parameters obtained for
the considered U, Pu, Cm, and Cf nuclei as functions of the neutron number $N$. 
Most of the derived parameters appear to stay nearly constant with $N$, and their
values also do not significantly differ between different isotopic
chains. This is conceptually very satisfying as it indicates the consistency of 
the approach and its predictive power. 
An exception is perhaps the $f$-boson energy $\epsilon_{f}$
depicted in panel (b) of Fig.~\ref{fig:parameter}, which exhibits an abrupt
structural change between neighboring isotopes. The rapid decrease of the quantity
$-\epsilon_{f}$ from $N=130$ to 136 indicates the development of
the octupole collectivity. 

%-----------------------------------------------------------------------
%
%	EXCITATION SPECTRA FOR EVEN-EVEN NUCLEI
%
%-----------------------------------------------------------------------
\begin{figure}[htb!]
\begin{center}
\includegraphics[width=\linewidth]{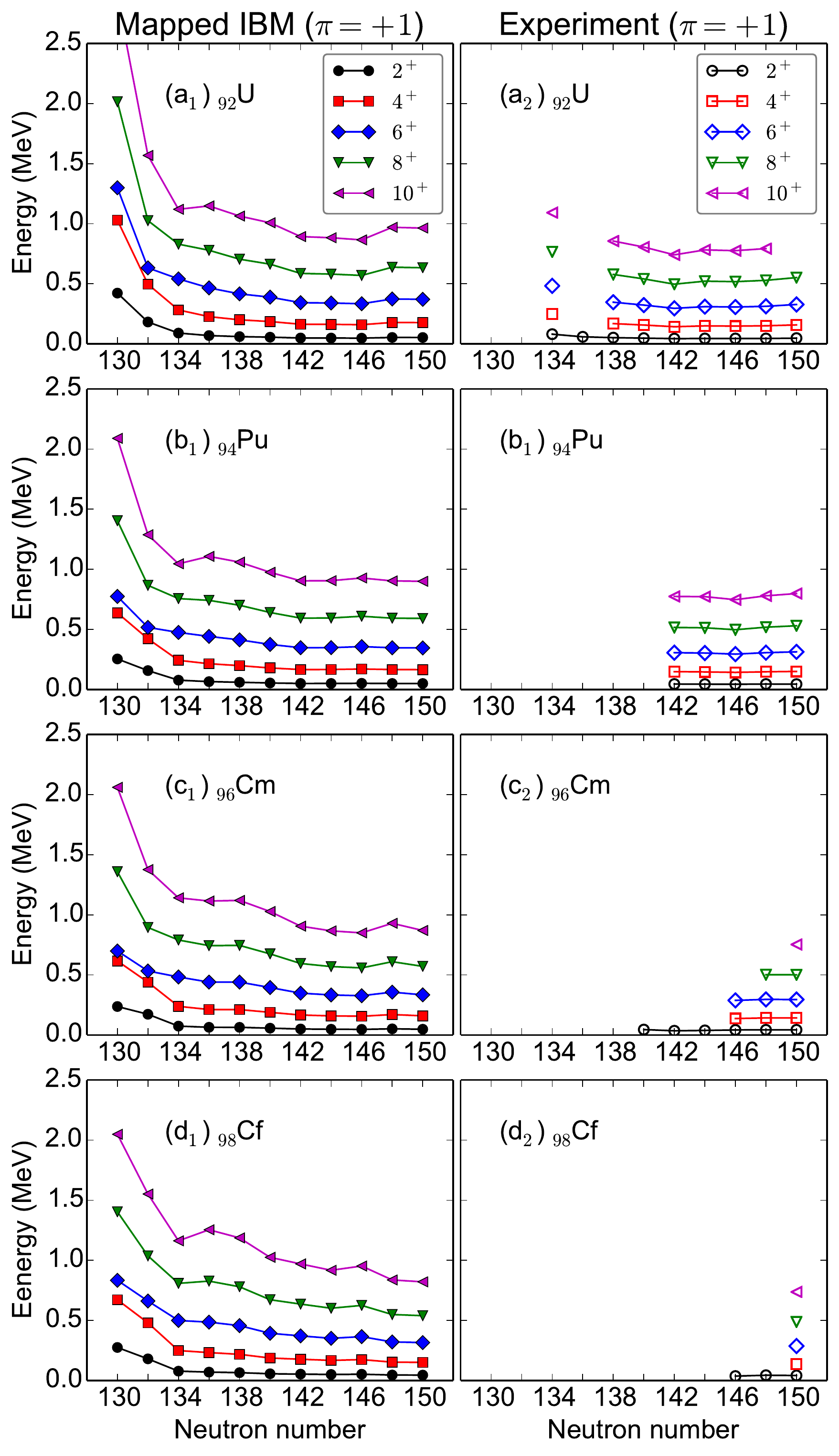}
\caption{Low-energy excitation spectra of positive-parity even-spin
 yrast states of $^{222-242}$U, $^{224-244}$Pu, $^{226-246}$Cm, and
 $^{228-248}$Cf computed by the diagonalization of the mapped $sdf$-IBM
 Hamiltonian  Eq.~(\ref{eq:ham}). Experimental data are taken from
 Ref.~\cite{data}. Results for Ra and Th isotopes can be found in Ref.~\cite{nomura2020oct}.} 
\label{fig:level-pos}
\end{center}
\end{figure}

%-----------------------------------------------------------------------
%
%	EXCITATION SPECTRA FOR EVEN-EVEN NUCLEI
%
%-----------------------------------------------------------------------
\begin{figure}[htb!]
\begin{center}
\includegraphics[width=\linewidth]{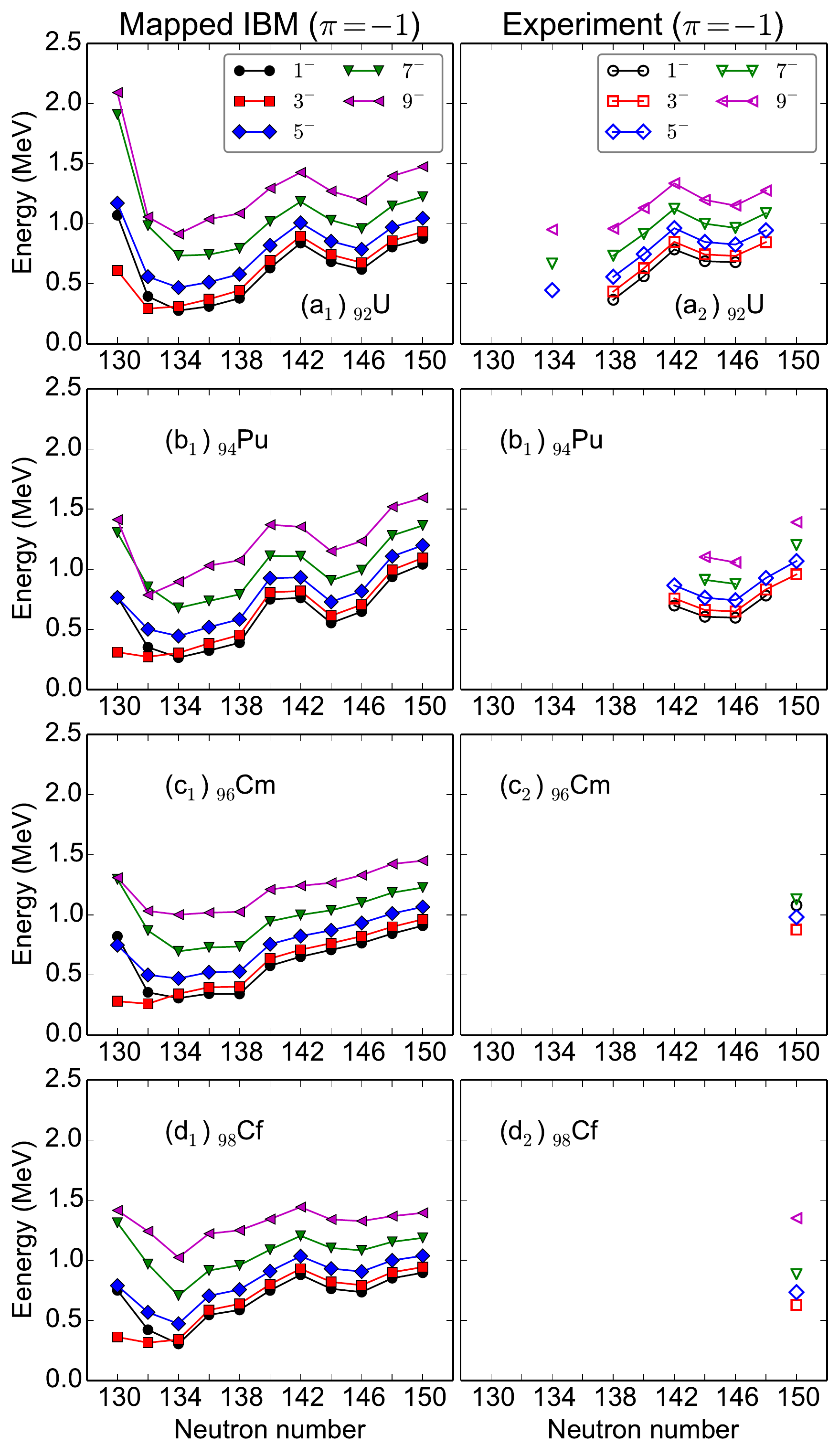}
\caption{Same as the caption of Fig.~\ref{fig:level-pos}, but for the
 odd-spin negative-parity yrast states.} 
\label{fig:level-neg}
\end{center}
\end{figure}

% Subsection:                 Evolution of low-energy excitation spectra

\subsection{Evolution of low-energy excitation spectra}

The excitation spectra for the low-lying even-spin
positive-parity and odd-spin negative-parity yrast
states of 
$^{222-242}$U, $^{224-244}$Pu, $^{226-246}$Cm, and $^{228-248}$Cf are
depicted in Figs.~\ref{fig:level-pos} and \ref{fig:level-neg} as
functions of $N$, respectively. 
In Fig.~\ref{fig:level-pos}, the calculated positive-parity levels in each
isotopic chain (panels (a$_1$) to (d$_1$)) are seen to 
decrease from $N=130$ to 134, as the quadrupole collectivity develops. 
A typical rotational band structure starts to appear from $N\approx
132$, and there is no significant change in positive-parity states from
$N=134$ on.  
Agreement between the theoretical and experimental (panels from (a$_2$)
to (d$_2$)) positive-parity 
levels is remarkable. It should be noted that 
the predicted level structure of the transitional nuclei with $N=130$
and 132 looks rather irregular: energy levels of the $4^+_{1}$ and 
$6^+_{1}$ are close to each other, at variance with the well-known vibrational
or rotational band patterns. 
As we show in Sec.~\ref{sec:nf}, such a irregularity in
the calculated levels seems to occur due to strong mixing
between positive- and negative-parity boson configurations in the
low-spin states of these nuclei.

The systematic behavior of the calculated negative-parity states is
depicted in Fig.~\ref{fig:level-neg} (panels (a$_1$) to (d$_1$)). 
The comparison of our results with the available experimental data
 (panels from (a$_2$) to (d$_2$)) is reasonable. 
For each considered isotopic chain, the predicted  negative-parity
states exhibit an 
approximate parabolic behavior centered around $N=134$ which corresponds with 
the nuclei where the
octupole minimum is  most pronounced in the SCMF-PESs. From $N=134$ on,
the negative-parity level energies keep increasing up to $N=142$, 
where another parabolic-like behavior sets in the U, Pu and Cf isotopes. 
The absence of permanent octupole deformation in those isotopes hints to
an increasing dominant role of dynamic octupole correlations.

%-----------------------------------------------------------------------
%
%	F-BOSON CONTENT
%
%-----------------------------------------------------------------------
\begin{figure}[htb!]
\begin{center}
\includegraphics[width=\linewidth]{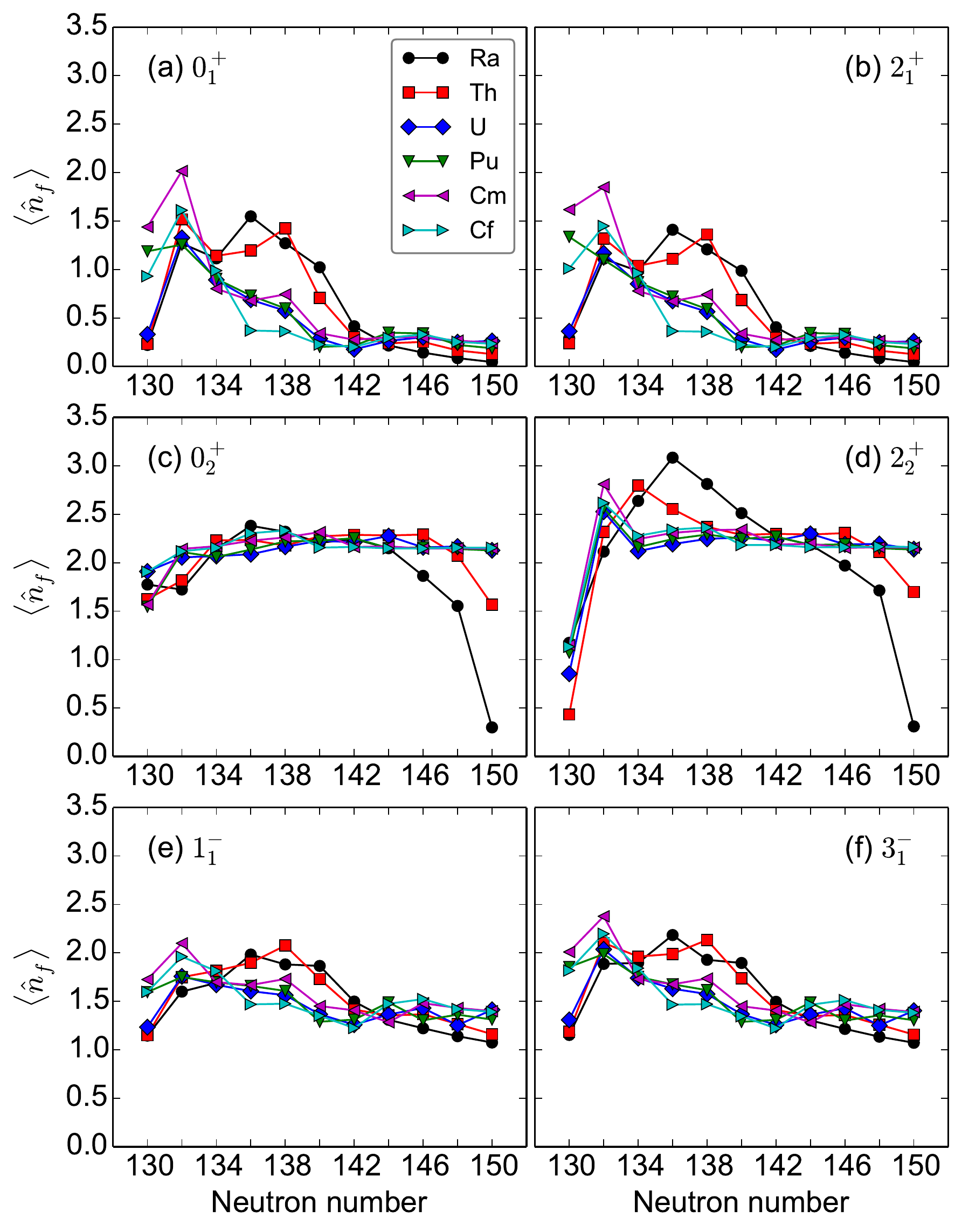}
\caption{Expectation values of the $f$-boson number operator
 $\braket{\hat{n_{f}}}$ in the IBM wave functions of the states 
$0^+_1$, $2^{+}_{1}$, $0^{+}_{2}$, $2^{+}_{2}$
, $1^{-}_{1}$, and $3^{-}_{1}$ of the Ra, Th, U, Pu, Cm, and Cf nuclei,
 plotted as functions of the neutron number. The values for the Ra and
 Th isotopes have been taken from \cite{nomura2020oct}. 
} 
\label{fig:nf}
\end{center}
\end{figure}

% Subsection:                  f$-boson contribution to low-lying states

\subsection{$f$-boson contribution to low-lying states\label{sec:nf}}

Next, we discuss the
contribution of the $f$ boson to the wave functions of the low-lying 
positive and negative parity states.  Figure~\ref{fig:nf} displays the 
expectation values of the $f$-boson
number operator $\braket{\hat{n}_{f}}$ computed in the IBM wave functions of
the states $0^{+}_{1}$, $2^{+}_{1}$, $0^{+}_{2}$, $2^{+}_{2}$,
$1^{-}_{1}$, and $3^{-}_{1}$. 
From panels (a) and (b) of Fig.~\ref{fig:nf}, we learn that the expectation value
$\braket{\hat{n}_{f}}$ is rather large in the transitional region
$132\leqslant N\leqslant 138$ indicating that the mixing between the $s-d$ and $f$
boson spaces is significant in the ground-state band. 
There is a marked difference in the results between the Ra and Th isotopes and the U,
Pu, Cm, and Cf isotopes with the neutron numbers 
$136\leqslant N\leqslant 140$: the $f$-boson
contributions to the ground-state bands of the U, Pu, Cm, and Cf
isotopes are by a factor two to five smaller than for the Ra and Th nuclei. 
Thus it follows that, in the present 
framework, the quadrupole-octupole
correlations become weaker as the proton number increases. 
For $N$ larger than 140, both the $0^+_1$ and $2^+_1$ states are made of the
$s$ and $d$ bosons alone, as the expectation values
$\braket{\hat{n}_{f}}\approx 0$. 

By looking at panels (c) and (d) in Fig.~\ref{fig:nf}, we conclude that 
the structure of the  wave functions of the $0^+_2$ and $2^+_2$ states
corresponds to a double-octupole phonon structure as the expectation
value $\braket{\hat{n}_{f}}\approx 2$. 
Empirical studies have interpreted that low-energy $K=0^+$ excited 
bands in the actinide region are partly accounted for by the coupling between
double octupole phonons \cite{zamfir2003,spieker2018}. 
Both isotopic and isotonic dependencies of these values are not as
strong as in the cases of the $0^{+}_{1}$ and $2^{+}_{1}$ states. 

In panels (e) and (f) of Fig.~\ref{fig:nf} we notice that for
the $1^-_1$ and $3^-_1$ states
$1.5\leqslant \braket{\hat{n}_{f}} \leqslant 2.0$ when
$N<140$. For larger values of $N$, $\braket{\hat{n}_{f}}$ gradually decreases to reach 
the value of one at $N\approx 150$. 
Therefore, we conclude that more than one $f$ boson is needed for
the employed EDF-to-IBM mapping procedure to reasonably describe
excitation spectra of low-lying negative-parity yrast states in actinide
nuclei with 
$N\leqslant 140$, while for heavier actinides, inclusion of only one
$f$ boson seems to suffice. 

%Note that results for those nuclei with $A\geqslant 242$ are not
%included in the plot. 
%The reason is that, since only one $f$ boson has been considered for
%these nuclei,  all the positive- and negative-parity
%states obviously have integer expectation values $\braket{\hat{n}_{f}}=0$ and 1,
%respectively. 

%-----------------------------------------------------------------------
%
%	Alternating parity band
%
%-----------------------------------------------------------------------
\begin{figure}[htb!]
\begin{center}
\includegraphics[width=\linewidth]{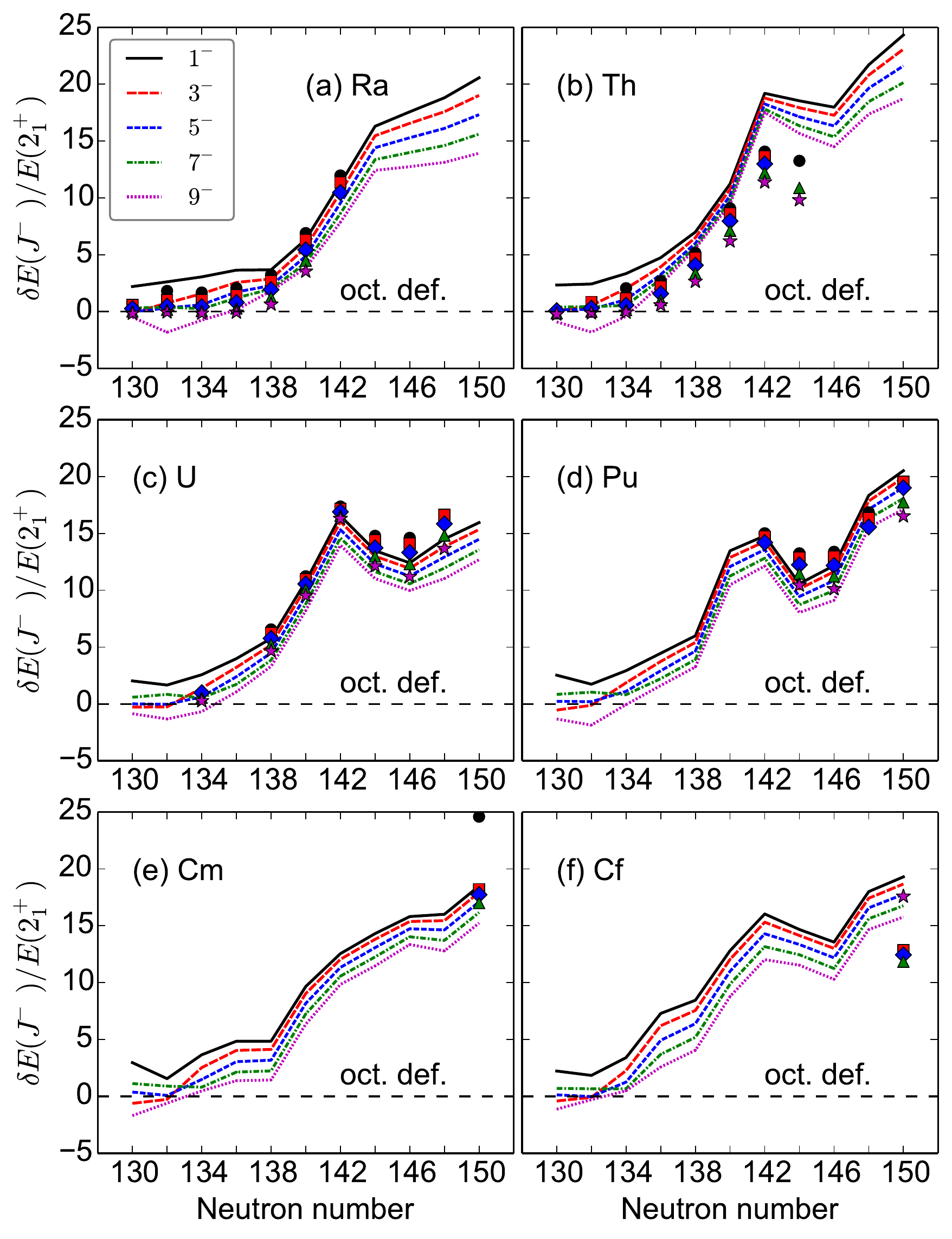}
\caption{The energy displacement $\delta E(J^{-})$ [defined in Eq.(\ref{eq:de})], normalized with respect to 
 the excitation energy of the  $2^{+}_{1}$ state, is shown as a function of 
 the neutron number. The theoretical values are connected by lines. The
 corresponding experimental \cite{data} values for the $J^-=1^{-}$, $3^{-}$, 
 $5^{-}$, $7^{-}$, and $9^{-}$ yrast states are represented by the solid
 circles, squares,  diamond, triangles, and stars, respectively. Results
 for the Ra and Th nuclei are from \cite{nomura2020oct}. The limit of
 stable octupole deformation $\delta 
 E(J^{-})=0$ is indicated in each panel by a broken horizontal line.}
\label{fig:de}
\end{center}
\end{figure}

% Subsection:                 Possible alternating-parity band structure

\subsection{Possible alternating-parity band structure}

To distinguish if the members of rotational bands are octupole-deformed or octupole
vibrational states, it is convenient to consider the quantity
\begin{align}
\label{eq:de}
\delta E(J^{-}) = E(J^{-}) - \frac{E((J+1)^{+}) + E((J-1)^{+})}{2}, 
\end{align}
where $E(J^{-})$ and $E((J\pm 1)^{+})$ represent excitation energies of the odd-spin
negative-parity and even-spin positive-parity yrast states,
respectively. If the positive- and negative-parity bands share an octupole deformed band head they form an 
alternating-parity doublet and the quantity $\delta E(J^{-})$ should be equal to zero. The
deviation from the limit $\delta E(J^-)=0$ means that the positive- and negative-parity
bands form separate bands, and therefore an octupole vibrational structure
emerges. 
In Fig.~\ref{fig:de}, the calculated $\delta E(J^-)/E(2^{+}_{1})$ values for the $J=1^{-}_{1}$
to $9^{-}_{1}$ states are plotted as functions of $N$. 
Let us take as examples the results for the Ra and Th isotopic chains in
panels (a) and (b) of Fig.~\ref{fig:de}. The ratios $\delta E(J^{-})/E(2^+_1)$ for each spin
are close to zero for a number of Ra and Th isotopes with neutron numbers below 
$N\approx 138$, i.e., $^{218-226}$Ra and 
$^{220-228}$Th. For heavier isotopes with $N\geqslant 140$, $\delta
E(J^{-})/E(2^+_1)$ values turn to increase with $N$. 
Thus the octupole vibrational states characterized by the octupole-soft
potential appear. 
Essentially the same trend is observed for the U, Pu, Cm, and Cf isotopes.

%-----------------------------------------------------------------------
%
%	E(3-) and B(E3) systematics
%
%-----------------------------------------------------------------------
\begin{figure}[htb!]
\begin{center}
\includegraphics[width=\linewidth]{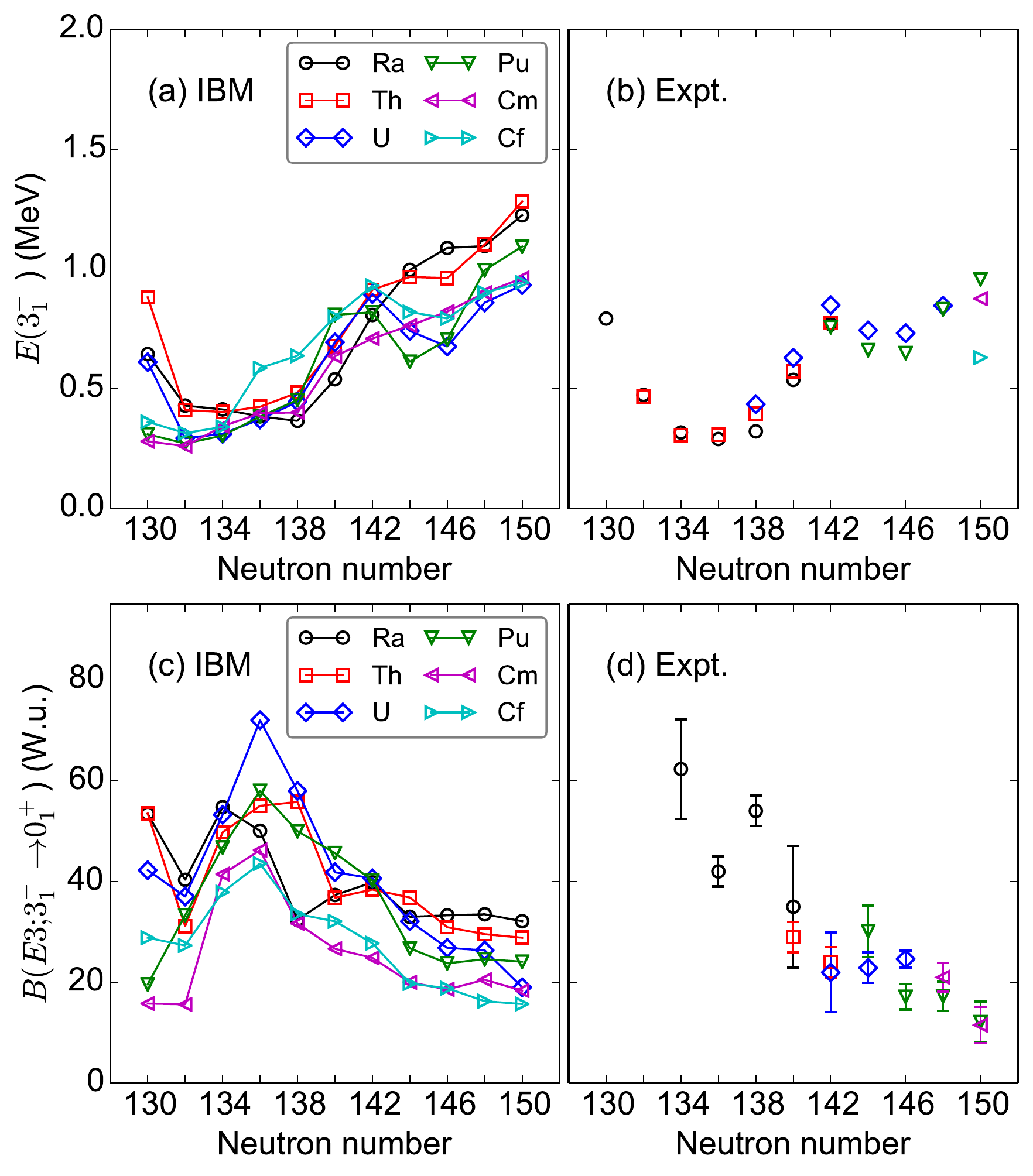}
\caption{Evolution of the theoretical and experimental excitation energy of the $3^-_1$
 state (panels (a) and (b)), and the $B(E3; 3^-_1\to 0^+_1$) transition strength in Weisskopf
 units (panels (c) and (d)) for the Ra, Th, U, Pu, Cm, and Cf isotopes as functions of
 the neutron number. Theoretical values for the Ra and Th nuclei
are taken from Ref.~\cite{nomura2020oct}. The experimental data are from
 Refs.~\cite{data,kibedi2002}
} 
\label{fig:e3}
\end{center}
\end{figure}

%-----------------------------------------------------------------------
%
%	E(E2) and B(E1) systematics
%
%-----------------------------------------------------------------------
\begin{figure}[htb!]
\begin{center}
\includegraphics[width=\linewidth]{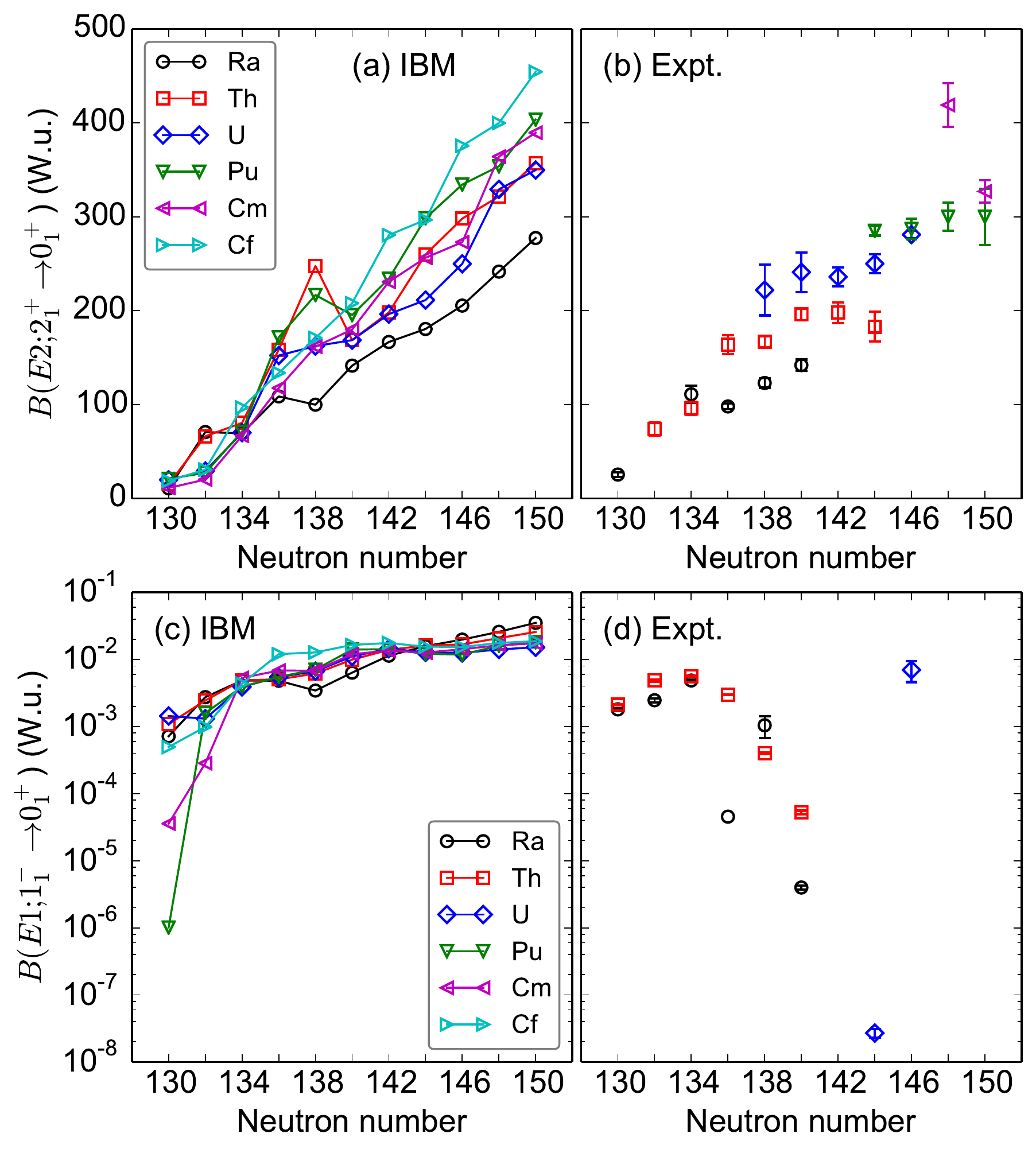}
\caption{Evolution of the theoretical and experimental $B(E2; 2^+_1\to
 0^+_1$) (panels (a) and (b)) and $B(E1; 1^-_1\to
 0^+_1$) (panels (c) and (d)) transition strength in Weisskopf
 units for the Ra, Th, U, Pu, Cm, and Cf isotopes as functions of
 the neutron number. Theoretical values for the Ra and Th nuclei
are taken from Ref.~\cite{nomura2020oct}. The experimental data are from
 Refs.~\cite{data}
} 
\label{fig:e2}
\end{center}
\end{figure}

% Subsection:                           $B(E3)$ reduced transition rates

\subsection{Transition rates}

Transition probabilities are computed with the electric dipole,
quadrupole and octupole 
transition operators $\hat T^{E\lambda} = e_{\lambda}\hat Q_{\lambda}$
($\lambda=1,2,3$). 
As in Ref.~\cite{nomura2020oct}, the effective charge $e_{\lambda}$ for
$\lambda=2$ and 3 depends on the boson number $n$, 
and is determined so that the IBM's intrinsic quadrupole (octupole) moments
calculated at $\beta_{\lambda}=\beta_{\lambda,\mathrm{min}}$ coincide
with the SCMF ones. 
Formulas used to determine the values of $e_{\lambda}$ can be found in
Ref.~\cite{nomura2020oct}. In the present calculations, a
slight modification to the formulas given in \cite{nomura2020oct} has
been considered:  
for the Cm, and Cf isotopes, an overall factor of the $e_{2}$ charge 
has been re-scaled so that the
experimental $B(E2;2^+_1\to 0^+_1)$ values for $^{244,246}$Cm are reasonably
reproduced. 
The employed $e_{\lambda=2,3}$ values are also shown in panels (k) and (l) of 
Fig.~\ref{fig:parameter}. The E1 transition operator is defined in
\cite{nomura2020oct}, and the same value of the E1 charge $e_{1}=0.0277$
$e$b$^{1/2}$ is used for all the considered isotopic chains. 
The $B(E3;3^{-}_{1}\to 0^{+}_{1})$ transition rates as well as the
excitation energies of the $3^{-}_{1}$ state are depicted as functions
of $N$ in Fig.~\ref{fig:e3}. The predicted $B(E3)$ values
show marked peaks at around $N=136$ with a maximum value of 72 W.u. for
$^{228}$U.  
This trend is correlated with the behaviors of the calculated and
experimental $E(3^{-}_{1})$ values plotted in
panels (a) and (b) of the same figure: the $B(E3;3^{-}_{1}\to
0^{+}_{1})$ values  
are inversely proportional to the $E(3^{-}_{1})$ ones.

In addition, we compare in Fig.~\ref{fig:e2} the
calculated and experimental $B(E2; 2^+_1\to 0^+_1$) and 
$B(E1; 1^-_1\to 0^+_1$) values. 
The predicted $B(E2)$ rates keep increasing with $N$ as the quadrupole
collectivity develops, and are in a good agreement with the data. 
However, the present model is unable to describe, even qualitatively, 
the empirical $B(E1)$ rates (panels (c) and (d) of Figs.~\ref{fig:e2}). 
This is mainly because of the fact that the E1 properties are less
collective in nature, while the IBM framework is built on correlated
pairs and deals with purely collective states. 
Another reason is that, within the $sdf$ boson space, the E1
transition operator only contains a term that is proportional to 
$(d^{\+}\tilde{f}+f^{\+}\tilde{d})^{(1)}$, and this simplified form may
not satisfactorily describe the details of the observed $B(E1)$
systematic. Inclusion of higher-order terms in the E1 operator or taking
into account explicitly the dipole $p$ boson with $J=1^{-}$ could
improve description of the E1 rates, 
but these extensions are clearly out of the scope of the present paper.

%-----------------------------------------------------------------------
%
%	Effective bet2 and bet3
%
%-----------------------------------------------------------------------
\begin{figure}[htb!]
\begin{center}
\includegraphics[width=\linewidth]{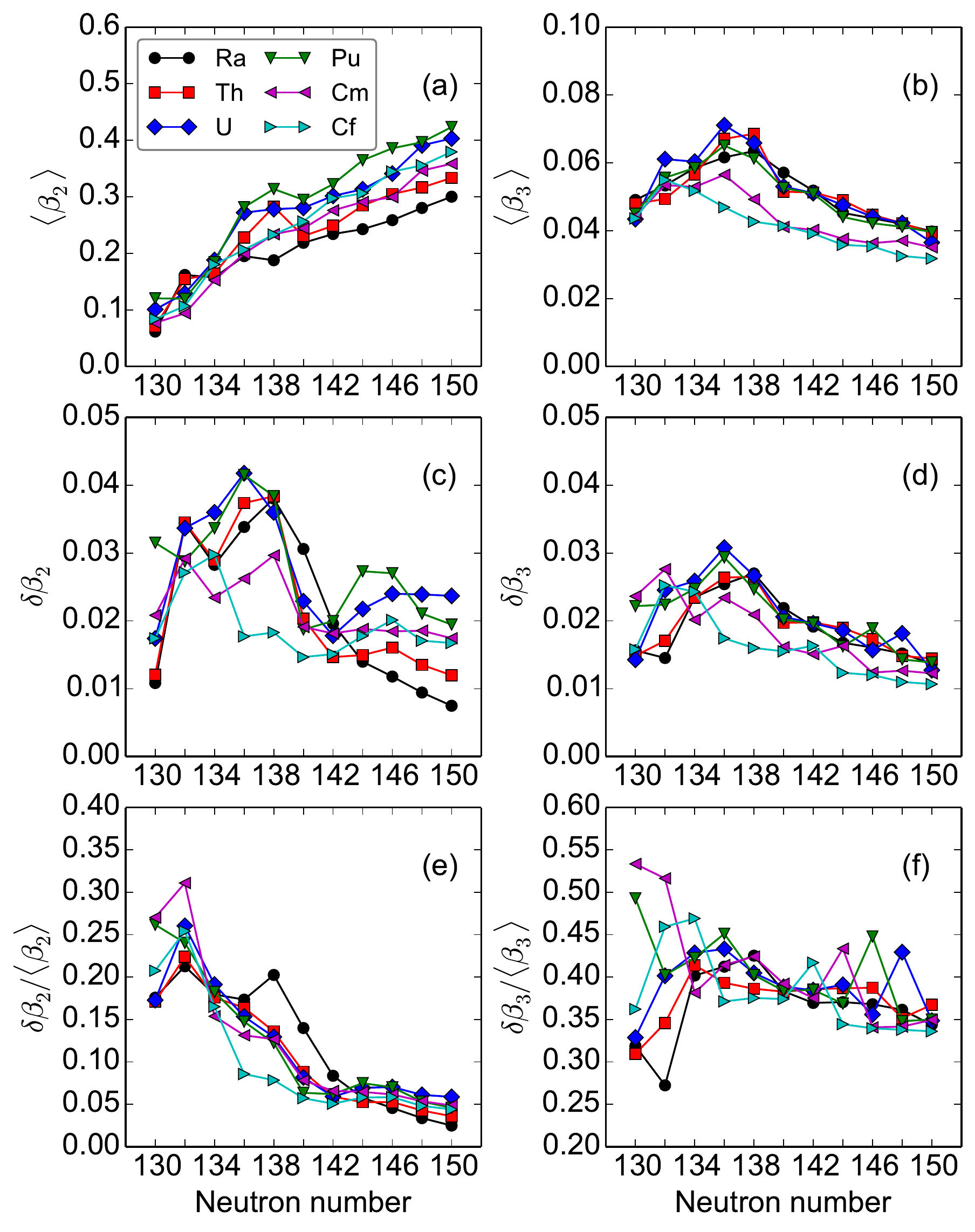}
\caption{Effective quadrupole $\braket{\beta_{2}}$ (a)
and octupole $\braket{\beta_{3}}$ (b) deformation parameters, variance $\delta\beta_{\lambda}$ (c,d), and the
 fluctuations $\delta\beta_{\lambda}/\braket{\beta_{\lambda}}$ (e,f) for
 $^{218-238}$Ra, 
 $^{220-240}$Th, $^{222-242}$U, $^{224-244}$Pu, $^{226-246}$Cm, and
 $^{228-248}$Cf, as functions of the neutron number. Values for the Ra
 and Th nuclei have been calculated based on the results from
 Ref.~\cite{nomura2020oct}. See the main text
 for details.}
\label{fig:bet23}
\end{center}
\end{figure}

% Subsection:             Effective quadrupole and octupole deformations

\subsection{Effective quadrupole and octupole deformations}

We further consider quadrupole and octupole shape invariants
\cite{cline1986,werner2000} calculated in the IBM ground state
$\ket{0^{+}_{1}}$: 
\begin{align}
%&q_{2}^{(\lambda)}=\bra{0^+_1}\bigl(\hat{T}^{E\lambda}\cdot\hat{T}^{E\lambda}\bigr)\ket{0^+_1}
%\\
%&q_{4}^{(\lambda)}=\braket{0^{+}_{1} | \bigl(\hat{T}^{E\lambda}\cdot\hat{T}^{E\lambda}\bigr)\bigl(\hat{T}^{E\lambda}\cdot\hat{T}^{E\lambda}\bigr) | 0^{+}_{1}}\\
\label{eq:q2}
q_{2}^{(\lambda)}=\sum_{i}
&(-1)^{J}
\braket{0^+_1     \| \hat{T}^{E\lambda} \| J^{\pi}_i}
\braket{J^{\pi}_i \| \hat{T}^{E\lambda} \| 0^+_1}\\ 
\label{eq:q4}
q_{4}^{(\lambda)}=\sum_{i,j,k}
&\braket{0^+_1     \| \hat{T}^{E\lambda} \| J^{\pi}_i}
\braket{J^{\pi}_i \| \hat{T}^{E\lambda} \| 0^+_j}
\nonumber \\
%&\quad\quad
&\times
\braket{0^+_j     \| \hat{T}^{E\lambda} \| J^{\pi}_k}
\braket{J^{\pi}_k \| \hat{T}^{E\lambda} \| 0^+_1}
\end{align}
where  $\braket{0^{+} \| \hat{T}^{E\lambda} \| J^{\pi}}$'s are reduced matrix
elements of the $E\lambda$ transition operators, and $J^{\pi}=2^{+}$ and $3^{-}$ for $\lambda=2$ and 3,
respectively. The sums in Eqs.~(\ref{eq:q2}) and (\ref{eq:q4}) include up
to 10 lowest $0^+$, $2^+$, and $3^-$ states. 
We can define the effective quadrupole and octupole deformation parameters
\begin{align}
\label{eq:beta}
\braket{\beta_{\lambda}} = \sqrt{\braket{\beta^{2}_{\lambda}}}
%= \frac{4\pi}{3eZR_{0}^{\lambda}}\sqrt{q_{2}^{(\lambda)}}
\end{align}
and the variance
\begin{align}
 \delta\beta_{\lambda}=\sqrt{\braket{\beta^{4}_{\lambda}}-\braket{\beta_{\lambda}^{2}}^{2}}/2\braket{\beta_{\lambda}},
\end{align}
where $\braket{\beta^{2m}_{\lambda}} =
({4\pi}/(3eZR_{0}^{\lambda}))^{2m}q_{2m}^{(\lambda)}$ ($m=1,2$). 

In Fig.~\ref{fig:bet23} we display the above mentioned quantities 
as functions of $N$: in panels (a) and (b) we show
$\braket{\beta_{\lambda}}$ for $\lambda=2$ and 3, respectively. 
In panels (c) and (d) the $\delta{\beta_{\lambda}}$ are displayed. Finally,
the fluctuations 
$\delta{\beta_{\lambda}}/\braket{\beta_{\lambda}}$ are presented in panels (e) and (f). 
The effective quadrupole deformation $\braket{\beta_{2}}$ increases
monotonously with $N$ as the quadrupole collectivity develops. 
Behavior of the effective octupole deformation parameter
$\braket{\beta_{3}}$ for each isotopic chain is characterized by a
parabolic trend with a marked peak at $N\approx 136$. 
The behavior of these quantities with neutron number is in agreement 
with the one of the intrinsic deformation
parameters shown in Fig. \ref{fig:eoct}. 
The variance for both the quadrupole and octupole deformations 
is large from $N\approx 132$ to $N\approx 140$. In such
transitional regions, potentials are soft in both $\beta_{2}$ and
$\beta_{3}$ (cf. Figs.~\ref{fig:pes-pu} and \ref{fig:pes-cf}),
and large shape fluctuations are present. 
This is  confirmed by rapid changes of the quantities
$\delta{\beta_{2}}/\braket{\beta_{2}}$ and
$\delta{\beta_{3}}/\braket{\beta_{3}}$ between $N\approx 132$ and
$N\approx 138$ (cf. panels (e) and (f) of Fig.~\ref{fig:bet23}). 

%We note that, in panels (c) and (d) of Fig.~\ref{fig:bet23}, the variance
%$\delta\beta_{\lambda}$ is nearly vanishing for those nuclei with $A>240$. 
%This is a direct consequence of the too small $q_{4}^{(\lambda)}$ values for these
%nuclei, as the reduced matrix elements between intermediate states, 
%$\braket{0^{+}_{j} \| \hat{T}^{E\lambda} \| J^{\pi}_{}}$ $(j>1)$, 
%are calculated to be substantially small in magnitude
%(cf. Eq.~(\ref{eq:q4})). This effect is probably 
%related to the fact that, in the present calculation,  only one $f$
%boson is considered for the $A>240$ nuclei. 

%-----------------------------------------------------------------------
%
%	Pu-240
%
%-----------------------------------------------------------------------
\begin{figure}[htb!]
\begin{center}
\includegraphics[width=\linewidth]{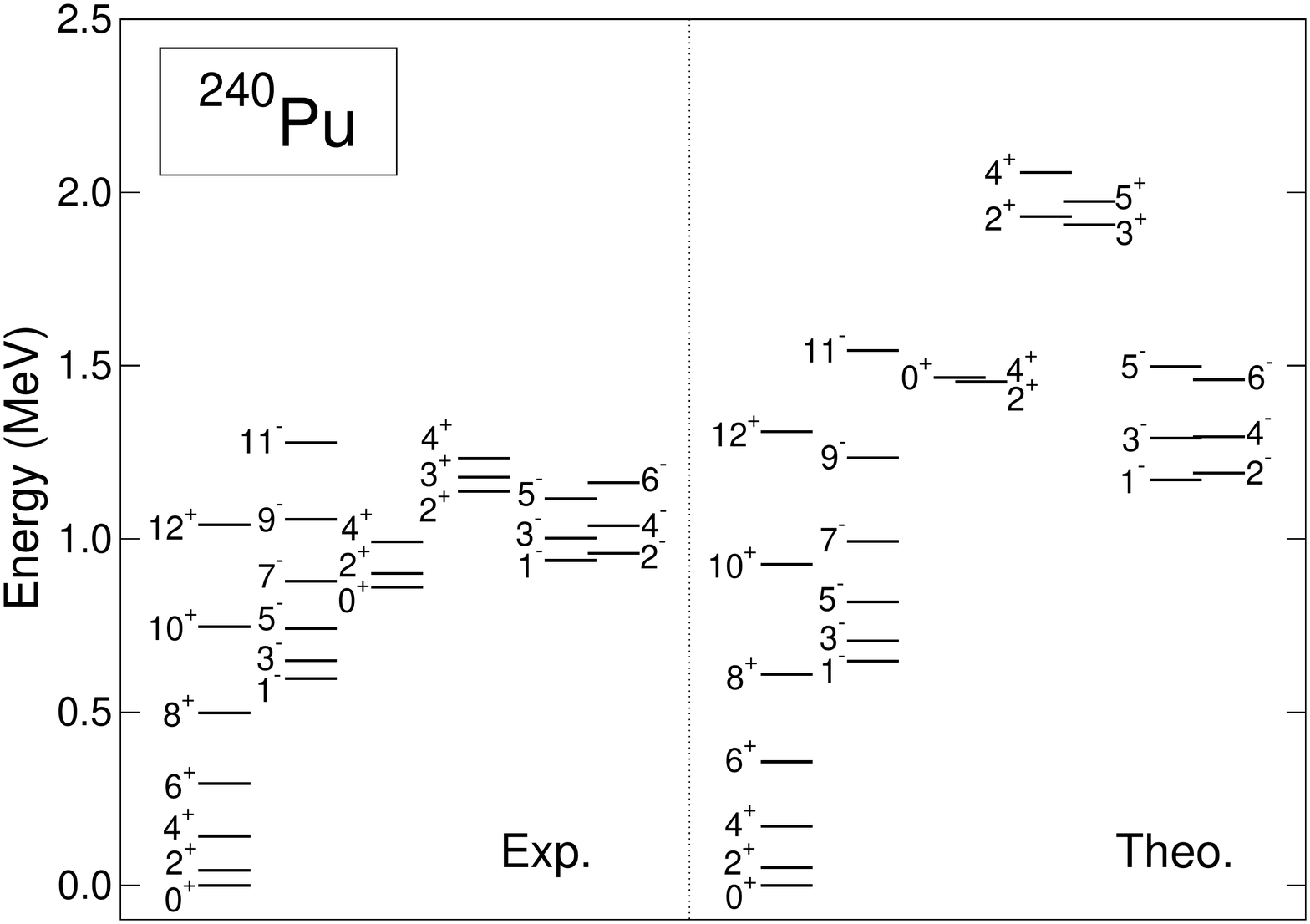}
\caption{Comparison of theoretical and experimental low-energy level
 schemes of $^{240}$Pu.}
\label{fig:pu240}
\end{center}
\end{figure}

% Subsection:                        Detailed level scheme of $^{240}$Pu

\subsection{Detailed level scheme of $^{240}$Pu}

% hereherehere

Finally, we examine the ability of the present  approach to
describe detailed spectroscopy in individual nuclei. 
As an illustrative example, in Fig.~\ref{fig:pu240} we compare 
theoretical and experimental low-energy level schemes of $^{240}$Pu. 
A good agreement between theory and experiment is observed in the 
lowest-lying members of the  positive- and
negative-parity rotational bands, as well as in the band built on the
$1^-_2$ state. 
The model, however, considerably overestimates the band-head energies of
the excited $K^{\pi}=0^+$ band built on the $0^+_2$ state and the
$\gamma$-vibrational ($K^{\pi}=2^{+}$) band on top of $2^+_{3}$. 
The discrepancies in the $K^{\pi}=0^{+}$ and $2^{+}$ bands are
most likely related to the fact that the underlying SCMF-PESs have too
steep potentials to be reproduced by the IBM ones. 
The description of the $\gamma$-band levels could be partly improved by
explicitly taking into account the triaxial degree of freedom in the
present calculation. It has been shown \cite{nomura2012tri} that a
specific form of three-body boson term plays an important role to lower
the $\gamma$ band. 
As we have seen in panels (c) and (d) of Fig.~\ref{fig:nf}, the IBM wave
functions of 
the $0^+_2$ state and the band built on it for most of the considered
actinide nuclei are mainly 
based on two-octupole-boson configurations. However, the fact that the
$0^+_2$ energy is overestimated 
may indicate a need for including additional building blocks in the IBM
framework to improve the agreement with experimental results.
For instance, inclusion of the dynamical pairing degree of freedom in
the IBM model space has been shown \cite{nomura2020pv} to have a
significant impact on the description of the excited $0^{+}$ states. 

We note that, except for the heavier isotopes Cm and Cf, a plenty of
spectroscopic data is available for the neighboring U and Pu nuclei as well. 
We have then carried out similar analyses, and confirmed that the level of
agreement between the predicted and experimental energy levels with both
parities for these nuclei is similar to $^{240}$Pu, while the same
problem as seen in $^{240}$Pu is commonly observed, that is, both the
excited $0^{+}$ and $2^{+}$ bands are considerably overestimated within
the mapped IBM framework. 

% ----------------------------------------------------------------------
%  Section:                                      Results and discussions
%
% ----------------------------------------------------------------------

\section{Summary\label{sec:summary}}

Octupole shapes and collective excitations in the actinide nuclei Ra,
Th, U, Pu, Cm, and Cf with  neutron number $130\leqslant N\leqslant 150$
have been investigated by using the EDF-based
IBM framework. Axially-symmetric quadrupole $\beta_{2}$ and octupole $\beta_{3}$
SCMF-PESs, obtained from constrained HFB calculations based on the
Gogny-D1M EDF, have been used to determine the
$sdf$-IBM Hamiltonian. Diagonalization of the mapped Hamiltonian
produces excitation spectra and transition strengths. 
The Gogny-D1M SCMF-PESs have suggested transitions from nearly
spherical ($N\approx 130$) to stable octupole-deformed ($N\approx 134$)
and to octupole-soft ($N\approx 138$) shapes along the considered U, Pu,
Cm, and Cf isotopic chains. 
Consistently with the empirical tendency, the calculated negative-parity
yrast states show a parabolic trend 
centered  at $N\approx 136$, and the $B(E3;3^-_1\to 0^+_1)$ transition rates
are predicted to take maximal values around $N\approx 136$, where the
SCMF-PESs exhibit the most pronounced octupole minima. 
%From the analysis of the quantity $\delta E(J^-)/E(2^+_1)$, 
The rotational bands of a number of nuclei exhibit the
alternating-parity pattern associated with a rigid octupole shape (cf. Fig.~\ref{fig:de}).  
%As we go away from the proton number $Z=88$ (Ra) to $Z=98$ (Cf), on the
%other hand, few isotopes show distinct octupole minimum.  
The effective $\beta_{2}$ and $\beta_{3}$ deformations and their variance
suggest large shape fluctuations near $N=134$. 
All the spectroscopic properties obtained from the mapped $sdf$-IBM
Hamiltonian exhibit tendencies that correlate with the variation of the
Gogny-D1M SCMF-PESs, and consistently suggest the onset of stable octupole
deformation around $N=136$ and that the transitions between
octupole-deformed and octupole-soft shapes occur systematically in
the actinide region. 
The spectroscopic results discussed in this paper also agree well, 
at least qualitatively, with the recent EDF-based spectroscopic
calculations \cite{xia2017,rayner2020oct} that covered the same region
of nuclei as the one studied here.

Even though our model allows for a detailed and economic description of
octupole-related spectroscopic properties, the current implementation of
the model is not able to describe quantitatively the spectra of non-yrast
states. For instance, the model has overestimated considerably the
excitation energies of the $0^+_{2}$ states and the $\gamma(K=2^+)$ band 
(cf. Fig.~\ref{fig:pu240}). This implies the necessity of including
those building blocks that are beyond the considered IBM framework, e.g.,
dynamical pairing and triaxial degrees of freedom (i.e., higher-order
terms in the IBM Hamiltonian). Since these new building
blocks have negligible contributions to the yrast states with both
parities, they would not alter the conclusion of the present work, that
is, the stable 
octupole shape occurs around $N\approx 134$ and octupole softness
emerges around $N\approx 138$ in actinide nuclei. 
Such extensions of the model will be required for a complete
spectroscopic study that involves an accurate description of excitation
energies and transition rates of non-yrast states. 
Another interesting topic is to extend the analysis
to odd-mass actinides. Work along these lines is in progress, and will
be reported in forthcoming articles.

\begin{acknowledgments}
This work has been supported by the Tenure Track Pilot Programme of 
the Croatian Science Foundation and the 
\'Ecole Polytechnique F\'ed\'erale de Lausanne, and 
the Project TTP-2018-07-3554 Exotic Nuclear Structure and Dynamics, 
with funds of the Croatian-Swiss Research Programme. The  work of LMR 
was supported by Spanish Ministry of Economy and Competitiveness (MINECO) 
Grant No. PGC2018-094583-B-I00.
This work has been partially supported by the Ministerio de Ciencia e
Innovaci\'on (Spain) under projects number PID2019-104002GB-C21, by the
Consejer\'{\i}a de Econom\'{\i}a, Conocimiento, Empresas y Universidad
de la Junta de Andaluc\'{\i}a (Spain) under Group FQM-370, by the
European Regional Development Fund (ERDF), ref.\ SOMM17/6105/UGR, and
by the European Commission, ref.\ H2020-INFRAIA-2014-2015
(ENSAR2). Resources supporting this work were provided by the CEAFMC
and the Universidad de Huelva High Performance Computer (HPC@UHU)
funded by ERDF/MINECO project UNHU-15CE-2848. 
\end{acknowledgments}

\bibliography{refs}

\end{document}